\documentclass[letterpaper, 10 pt, conference]{ieeeconf}  
\usepackage[noadjust]{cite}
\usepackage{filecontents}
\usepackage{cite}

\usepackage{epsfig}
\usepackage{graphicx}
\usepackage{amsmath}
\usepackage{amssymb}
\usepackage{mathrsfs}
\usepackage{url}

\usepackage{algorithm}
\usepackage{float}
\usepackage{subfiles}
\usepackage{nth}
\usepackage[noend]{algpseudocode}

\IEEEoverridecommandlockouts                              

\title{\LARGE \bf Gesture-based Piloting of an Aerial Robot using Monocular Vision
}

\author{Ting Sun$^{1}$, Shengyi Nie$^{1}$, Dit-Yan Yeung$^{2}$, Shaojie Shen$^{1}$
\thanks{$^{1}$ Department of Electronic \& Computer Engineering, Hong Kong University of Science and Technology, Hong Kong, China. {\tt\small (tsun, snie, eeshaojie)@ust.hk} }
\thanks{$^{2}$ Department of Computer Science, Hong Kong University of Science and Technology, Hong Kong, China. {\tt\small dyyeung@cse.ust.hk}}%
}

\begin{document}

\maketitle
\thispagestyle{empty}
\pagestyle{empty}

\begin{abstract}
Aerial robots are becoming popular among general public, and with the development of artificial intelligence (AI), there is a trend to equip aerial robots with a natural user interface (NUI).  Hand/arm gestures are an intuitive way to communicate for humans, and various research works have focused on controlling an aerial robot with natural gestures.  However, the techniques in this area are still far from mature.  Many issues in this area have been poorly addressed, such as the principles of choosing gestures from the design point of view,  hardware requirements from an economic point of view, considerations of data availability, and algorithm complexity from a practical perspective.  Our work focuses on building an economical monocular system particularly designed for gesture-based piloting of an aerial robot.  Natural arm gestures are mapped to rich target directions and convenient fine adjustment is achieved.  Practical piloting scenarios, hardware cost and algorithm applicability are jointly considered in our system design.  The entire system is successfully implemented in an aerial robot and various properties of the system are tested.

\end{abstract}

\section{INTRODUCTION}
\label{sec:introduction}
Nowadays aerial robots have moved beyond military applications and become popular in the civilian market. Various aerial robots are used to capture photos and videos from the aerial perspective. Users commonly control an aerial robot with a controller, and the control signal is transmitted through wireless communication channels.

With the development of artificial intelligence (AI), there is a trend to develop natural user interfaces (NUIs) \cite{neto2013real,goutsu2015gesture,fernandez2016natural,burke2015pantomimic} that facilitate the interaction between humans and machines. Hand/arm gestures are a natural and intuitive means of expression for humans, and using such gestures would allow us to instinctively command an aerial robot, in a way similar to commanding our pets. Thus, gesture-based human-computer interaction (HCI) is a promising research area for aerial robot control.

Existing gesture-based control systems of aerial robots are quite different from each other \cite{urban2004recognition,lementec2004recognition,ng2011collocated,sanna2013kinect,higuchi2013flying,pfeil2013exploring,naseer2013followme}. From a design point of view, different gesture vocabularies are mapped to different chosen commands.  Most designs treat gesture recognition as a classification problem, where a few gestures are recognized and map to arbitrary chosen commands.  However, an aerial robot can perform various functions, so rather than arbitrarily picking some of them, it is possible to focus on a particular scenario, enabling a tailored design.  E.g. to design a gesture to trigger a ``taking a picture" action, we need to consider that the system may need to handle a scene with multiple people close together and decide a proper delay between receiving the command and taking the action; to design gestures to demonstrate a set of special trajectories, a set of movements that mimic these trajectories may be a good choice; and to design gestures to command the aerial robot to take off or land, we need to consider gestures sophisticated enough to ensure that they won't be triggered by accident.  A good NUI is application dependent, and simply involving gestures does not guarantee intuitive human-machine interaction and enjoyment.  

In this work, we propose a monocular gesture-based control system,  which is particularly designed to pilot an aerial robot.  Despite its physical speed limitations, a small quadrotor, which is the most common aerial robot, can travel in any direction in open space, and it is desirable for the commands generated by our system to cover as many directions as possible.  Our design jointly considers this piloting scenario, the natural gestures of humans, the aerial robot's behaviors expected by the user,  and the command transition between different directions targeted by the aerial robot.  Our design not only realizes a pet-like enjoyable interaction on aerial robots, but also helps the users to get rid of the unwieldy controller in certain scenarios and has potential to be developed into some game.

Human gestures can be captured by expensive equipment, such as a data suit, data gloves or an optical motion capture system, but since aerial robots are commonly equipped with various cameras, capturing human gestures in a vision-based manner using these cameras saves additional hardware costs and payload.  The sensor required by our system is only a color camera, which is economical and applicable in both indoor and outdoor cases.  Based on the input video stream captured by an onboard color camera, our system recognizes a stretched out hand and command the aerial robot to fly in the direction pointed to by the user.  In contrast to modeling it as a classification system, directly using the directions pointed to by the user without quantization enables fine piloting in any direction parallel to the image plane, and it is also intuitive and easy for users to change their own gestures in order to adjust the behavior of the aerial robot.  A pair of gestures is designed to pilot in the direction perpendicular to the image plane.

As for recognition algorithms, NUI requires real-time processing and robustness. Thanks to the techniques of computer vision and artificial intelligence, machines are able to see and interpret vision information efficiently.  Our system involves a face detector, a tracker, a skin detector and a hand-crafted decision tree to generate commands, and the entire algorithm can easily run onboard in real time.  From the implementation point of view, similar works do not use the same platforms, so rather than a numerical comparison of performances, the complete implementation of the entire system is emphasized.

The remainder of this paper is organized as follows. Section~\ref{sec:related work} reviews previous works on gesture recognition and human-machine interaction. Our proposed system is presented in Section~\ref{sec:proposed system}, which is then followed by experimental results in Section~\ref{sec:experiments}. Section~\ref{sec:conclusion} concludes this paper.

\section{RELATED WORK}
\label{sec:related work}
There are many inspiring algorithms solely concerning vision-based gesture recognition \cite{li2011hand,kurakin2012real,ren2013robust,shotton2013real,liu2016real}, and various works have addressed the human-machine interaction issue \cite{boehme1997neural,waldherr1998neural,iba1999architecture,urban2004recognition,lementec2004recognition,mortezapoor2012multi,pfeil2013exploring,sanna2013kinect,hu2003visual,nickel20043d,hasanuzzaman2004real,ng2011collocated,higuchi2013flying,naseer2013followme,burke2015pantomimic}. The design philosophies, however, are usually just generally mentioned.  In \cite{ng2011collocated}, W.S. Ng and E. Sharlin particularly emphasizes that their design is inspired by falconeering, and it is evaluated by users.  Kevin P. Pfeil et al. in \cite{pfeil2013exploring} explore multiple sets of upper body 3D interaction metaphors, and the evaluation by users shows a preference toward intuitive and easy gestures.  Our command design aims at recognizing simple and intuitive gestures, and mapping them to as many piloting commands (i.e. target directions) as possible.  

Since a depth image highlights the foreground structure and has much lower dimensions than a color image, it is a popular input for gesture recognition algorithms. The depth is usually calculated from stereo images \cite{nickel20043d,hasanuzzaman2004real,li2011hand} or directly captured by an active sensor \cite{mortezapoor2012multi,naseer2013followme,pfeil2013exploring,ren2013robust,shotton2013real,sanna2013kinect,burke2015pantomimic}.  Kinect is one of the best-performing, and thus most popular depth sensors used in gesture recognition, but it only works in indoor environments, while piloting an aerial robot is a popular outdoor activity as well.  Our system only requires a color camera, which is very low-cost,  and can be applied in both the indoor and outdoor environments.  When color images are used in gesture recognition, skin detection is widely adopted to find the region of interest (i.e, hands and face) \cite{boehme1997neural,hu2003visual,nickel20043d,hasanuzzaman2004real,li2011hand,liu2016real}, and this is applied in our system as well. 

Static gestures are usually recognized by various hand-crafted features and rule-based algorithms \cite{boehme1997neural,hu2003visual,hasanuzzaman2004real,nickel20043d,li2011hand,kurakin2012real,shotton2013real,ren2013robust,burke2015pantomimic,liu2016real}, and dynamic gestures are typically recognized by state-space models \cite{iba1999architecture,nickel20043d,kurakin2012real}. A neural-networks-based method is used in \cite{waldherr1998neural} for image interpretation, predicting the angles of the arm from an image segment.  We focus on static gesture recognition and the recognition results of multiple frames are summarized together to robustly generate a command.  Our recognition algorithm is rule-based and is not data-driven.

\section{PROPOSED SYSTEM}
\label{sec:proposed system}
\begin{figure}
 \centering
 \vspace{0.2cm}
 \includegraphics[width=0.43\textwidth]{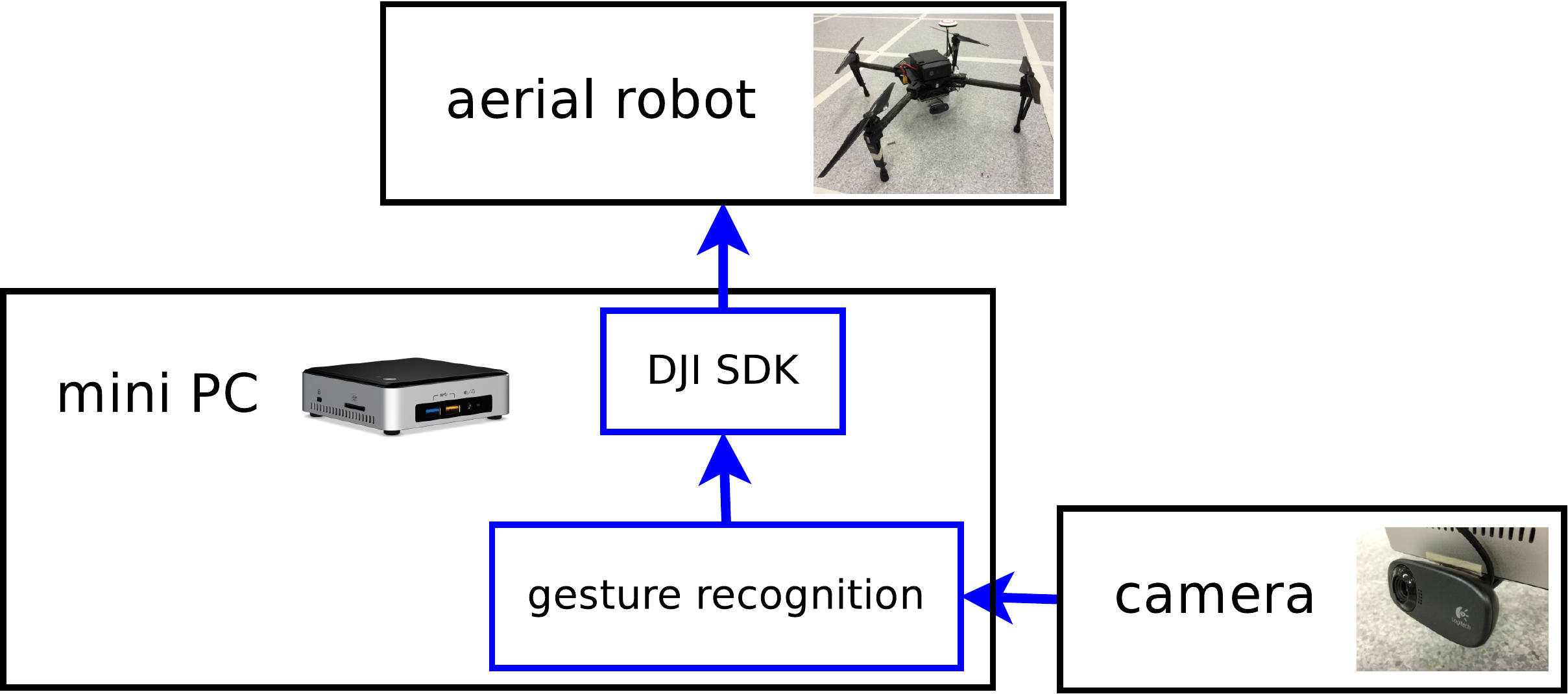}
 \caption{System overview. A black rectangle represents a piece of hardware and a blue rectangle is a software module. The communication between different parts is shown by blue arrows. All the pieces of hardware are onboard.}
 \label{fig:sys}
 \vspace{-1em}
\end{figure}

In this paper we propose a monocular gesture-based piloting system for aerial robots. The overview of the whole system is shown in Fig.~\ref{fig:sys}.  We focus on the command design and gesture recognition module, and in the following, we will first describe the design of our commands, followed by the detailed procedure of the gesture recognition algorithm.  

\subsection{Command Design}
\label{subsec:command design}
We assume that during the interaction, the user keeps visible for the aerial robot, and there are no other people close by.  Considering that the quality of the video captured by an onboard camera is usually not very high due to the distance between the user and the aerial robot, camera motion, uncontrollable outdoor lighting conditions etc, arm gesture instead of hand gesture, is recognized. 

As mentioned before, our piloting system maps natural arm gestures to as many target directions as possible. Here we use single arm pointing gestures and we do not distinguish between left and right arms.  We first detect both the user and his/her hand, which is not in a resting position, then command the aerial robot to fly in the direction pointed to by the user, with the velocity proportional to the length of the distance from the hand to the the center point between the shoulders.  This mapping results in a very intuitive interaction between the user and the aerial robot, and the user does not need to memorize specially designed gesture vocabularies in most cases.  

The position of the user's hand has 3D coordinates, but we can only extract its 2D projection on the image plane. Thus, we design a pair of special gestures to command the aerial robot to go further and come closer.  Both these commands are triggered when a hand is placed in front of the body.  If the hand raises above the body center, the system will command the aerial robot to move closer and if the hand is placed below the body center, the system will command the aerial robot to go further.  

\begin{figure*}
 \centering
 \vspace{0.15cm}
 \includegraphics[width=0.19\textwidth, height=0.18\textwidth]{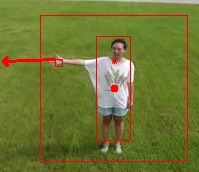}
 \includegraphics[width=0.19\textwidth, height=0.18\textwidth]{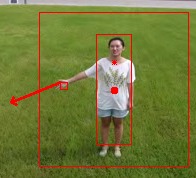}
 \includegraphics[width=0.19\textwidth, height=0.18\textwidth]{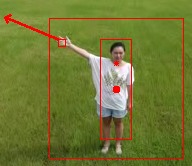} 
 \includegraphics[width=0.19\textwidth, height=0.18\textwidth]{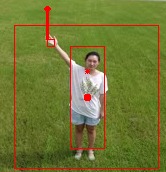} 
 \includegraphics[width=0.19\textwidth, height=0.18\textwidth]{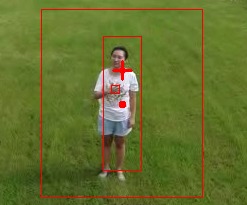}   \\
 \includegraphics[width=0.19\textwidth, height=0.18\textwidth]{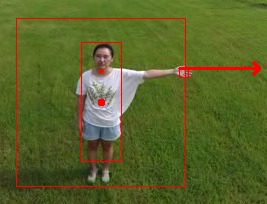}
 \includegraphics[width=0.19\textwidth, height=0.18\textwidth]{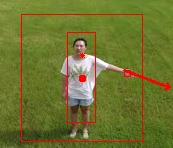}
 \includegraphics[width=0.19\textwidth, height=0.18\textwidth]{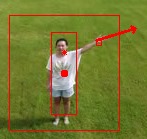}
 \includegraphics[width=0.19\textwidth, height=0.18\textwidth]{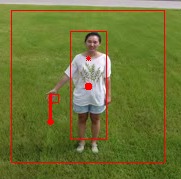}
 \includegraphics[width=0.19\textwidth, height=0.18\textwidth]{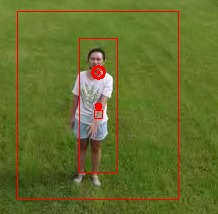}
 \caption{A few samples of the gestures we use.  The commands parallel to the image plane are shown by a red arrow, the ``come closer" command is shown by a cross in front of the upper body of the user and the ``go further" command is shown by a small circle.}
 \label{fig: gesture design}
 \vspace{-1em}
\end{figure*}

Theoretically, the user can point in any direction parallel to the 2D image plane, and in most situations during the interaction, the pointing direction of the hand indicates where the user wants the aerial robot to go.  However in practice, three cases of hand positions can not be interpreted naively in this way according to common sense: when the hand is resting at the side of the user, and when the hand is pointing up or pointing down. The resting position should not trigger aerial robot movement in any direction, and when the hand is raised high as shown in the fourth image in the top row of Fig.~\ref{fig: gesture design}, though not exactly pointing up in the center, the user probably wants the aerial robot to go straight up rather than sideling up,  and the same applies to the pointing down hand gesture, as shown in the fourth image in the bottom row of Fig.~\ref{fig: gesture design}.  We adjust the generated command in these three cases so that our system shows more reasonable behaviors.  

\begin{figure}
 \centering
 \includegraphics[width=0.45\textwidth]{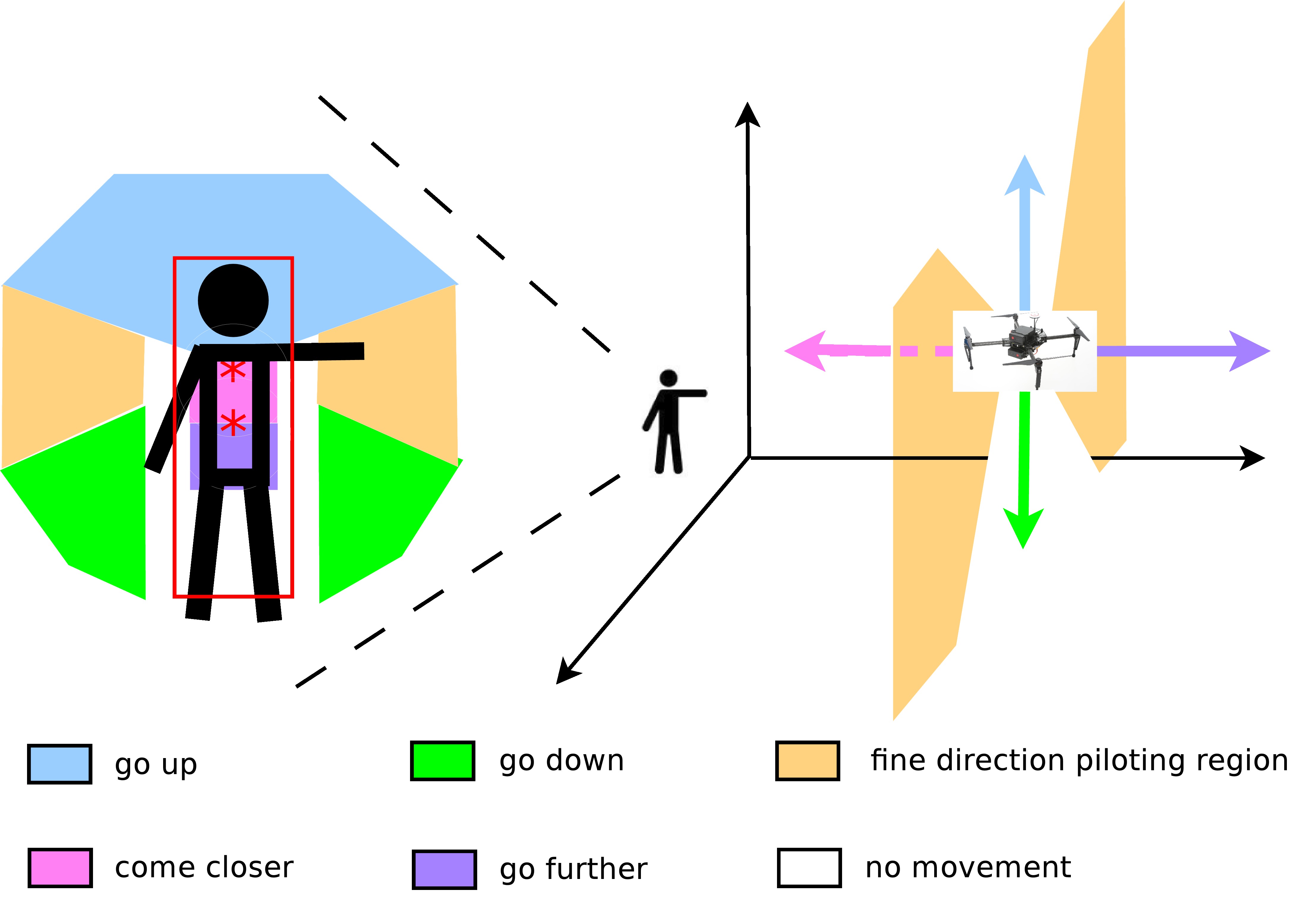}
 \caption{Command design summary.  The red bounding box locating the user is obtained by tracking (detailed in Subsection~\ref{subsubsec:dsst}).  When the user places a hand in different color regions in the 2D image, the system will command the aerial robot to travel in different directions colored correspondingly. ``Fine direction piloting region" means that when the user places a hand in this region, the system will command the aerial robot to travel in whatever direction is pointed to by the user without quantization.  ``No movement" means the user's hands are in a resting position, and no movement of the aerial robot is triggered.}
 \label{fig:command design}
 \vspace{-1em}
\end{figure}

Fig.~\ref{fig:command design} summarizes our command design.  As shown in the figure, when the user places a hand in different color regions in the 2D image, the system will command the aerial robot to travel in different directions colored correspondingly.  Note that in the orange region we do not quantize the direction obtained from the pointing gesture of the user, so in this region very fine piloting is achieved.  Fig.~\ref{fig: gesture design} shows some actual cases.  The user is performing piloting gesture, and the recognition results are marked in red on the images.  The commands parallel to the image plane are shown by a red arrow, the ``come closer" command is shown by a cross in front of the the center between the shoulders of the user and the ``go further" command is shown by a small circle.

Another strong advantage of our design is that in the fine piloting region, when the user observes that the aerial robot deviates from the desired path, he/she can easily correct the aerial robot by changing his/her own gesture intuitively. E.g. when the user sees the aerial robot is traveling higher than desired, he/she only needs to lower his/her arm, rather than recall a sophisticated gesture vocabulary to figure out what to do. 

\subsection{Gesture Recognition}
\label{subsec:gesture recognition}
\begin{figure}
 \centering
 \vspace{0.2cm}
 \includegraphics[width=0.43\textwidth]{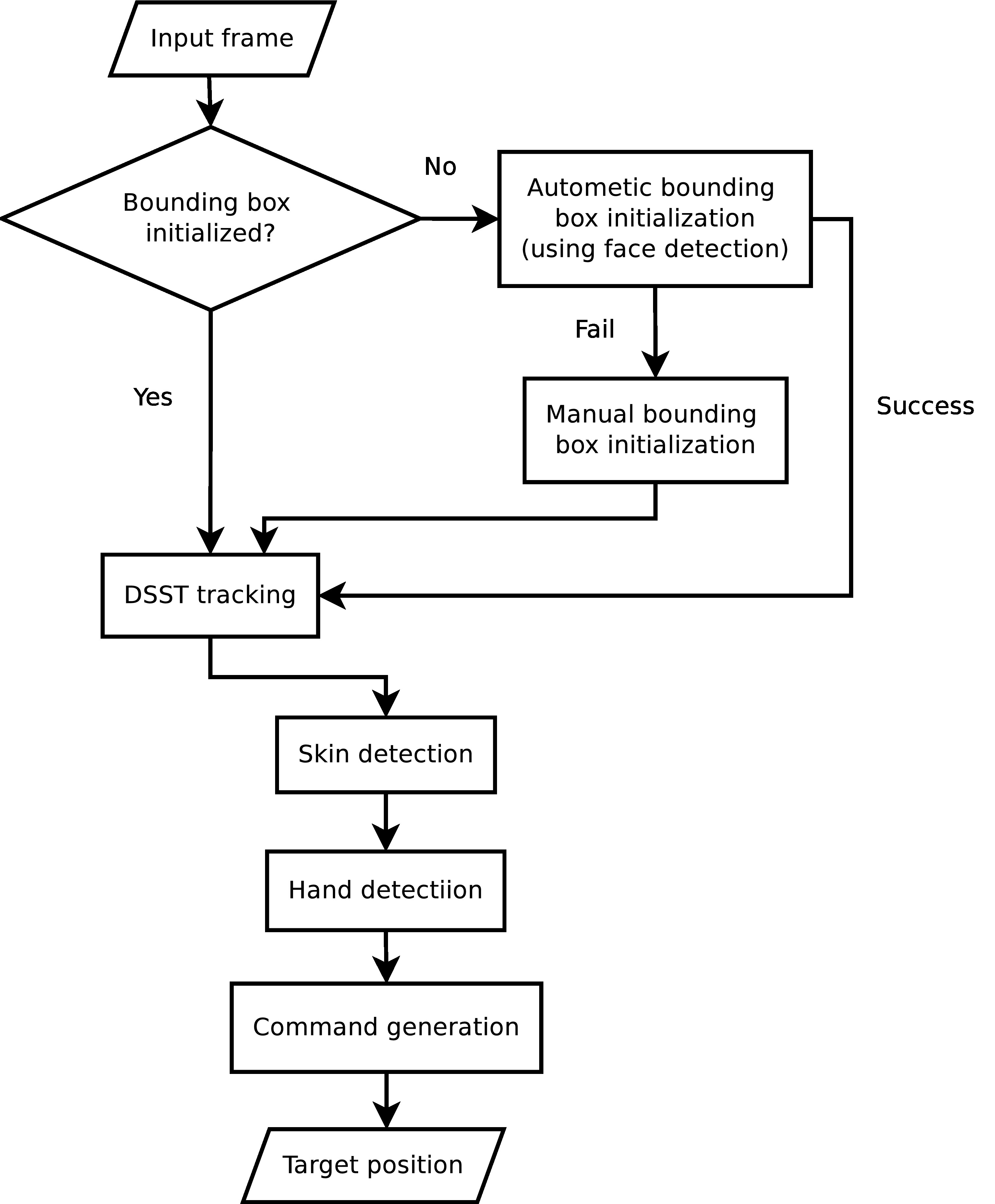}
 \caption{The flow chart of our gesture recognition method.}
 \label{fig:algorithm}
 \vspace{-0.5em}
\end{figure}

The flow chart of our gesture recognition method is shown in Fig.~\ref{fig:algorithm}.  The static gesture in each frame is recognized first, and the recognition results of multiple frames are summarized to robustly generate a piloting command.  The static gesture in a frame is represented by a 2D vector indicating a stretched out hand position relative to the center point between the shoulders of the user (the pointing gesture), or a hand position in front of the body relative to the center of the bounding box tightly containing the user.  This vector specifies the next target position of the aerial robot desired by the user.  In the recognition method, a discriminative scale space tracker (DSST) \cite{danelljan2014accurate} is applied to detect the user first, and then a non-resting hand is detected according to the results of both tracking and skin detection.  Typically a tracker is initialized by manually drawing a bounding box that tightly contains the target object. Considering that in our scenario the target is always a person, we adopt a face detector to automatically initialized the tracker, and manual initialization is also implemented in case the face detection fails.  The details of each major step are given as follows.

\subsubsection{Face Detection}
We use the Haar feature-based cascade classifier in OpenCV \cite{opencv_library,viola2001rapid,lienhart2002extended} for face detection.  This detector has two main characteristics. Firstly it uses very simple features like edges, lines and center-surround points \cite{lienhart2002extended} to detect an object, such as a face \cite{viola2001rapid}.  These simple features not only enable fast calculation, but are also very easy to scale.  The second characteristic is the cascade classifier, which is a concatenation of several simple classifiers.  A successful face sample should be able to pass all the simple classifiers at each stage \cite{lienhart2002extended}.  The result of the face detection is given by a bounding box tightly containing the face.  If a face is successfully detected in the first frame, a bounding box containing the whole body of the user is calculated from the face bounding box according to a preset head-body ratio.

\subsubsection{Discriminative Scale Space Tracker (DSST)}
\label{subsubsec:dsst}
The DSST \cite{danelljan2014accurate} extends the standard discriminative correlation filters to multidimensional features, and 3-dimensional filters are used for scale-space localization of the target. Considering a d-dimensional feature map representation of an image, let $f$ be a rectangular patch, and the $l$th dimension of $f$ is denoted by $f^l$.  In order to find an optimal correlation filter $h$ consisting of $h^l$ for each dimension, \cite{danelljan2014accurate} minimizes the following objective function:

\begin{equation}
 \epsilon = \left\| \sum_{l=1}^{d}h^l \ast f^l -g \right\|^2 + \lambda \sum_{l=1}^{d} \left\| h^l\right\|^2 
 \label{equ: filter objective}
\end{equation}

where $g$ is the desired correlation and $\lambda > 0$ weights the regularization term. The solution of (\ref{equ: filter objective}) in the Fourier domain is:

\begin{equation}
 H^{l} = \frac{\bar{G} F^l}{\sum_{k=1}^d \bar{F^k} F^k + \lambda} 
 \label{equ: fourier solution}
\end{equation}

Using $A_t^l$ and $B_t$ to denote the numerator and denominator in (\ref{equ: fourier solution}), in order to obtain a robust approximation, the update equation with learning rate parameter $\eta$ can be written in the following form:

\begin{equation}
 A_t^l = (1-\eta)A_t^l + \eta \bar{G_t} F_t^l 
\end{equation}

\begin{equation}
 B_t = (1-\eta)B_t + \eta \bar{G_t} \sum_{k=1}^d \bar{F^k} F^k 
\end{equation}

Given the trained filter, the score $y$ of a rectangular region $z$ is computed as

\begin{equation}
 y = \mathscr{F}^{-1} \left \{ \frac{\sum_{l=1}^d \bar{A^l Z^l}}{B+\lambda} \right \} 
\end{equation}

The new target state is the $z$ that has the maximum $y$. While searching the target in the current frame, DSST starts at the target location in the previous frame, and the 2-D translation subspace is searched first.  Then the 1-D scale subspace is determined. In our system, the tracking result is a bounding box $\mathbf{B}_{user}$ containing the target object, i.e, the person interacting with the aerial robot.

\subsubsection{Skin Detection}
\label{subsubsec:skin detection}
\begin{figure}
 \centering
 \includegraphics[width=0.11\textwidth]{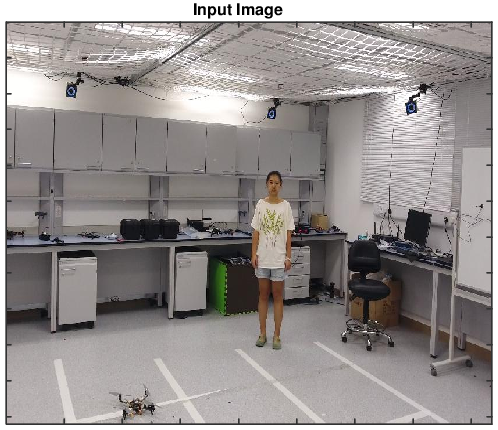}
 \includegraphics[width=0.11\textwidth]{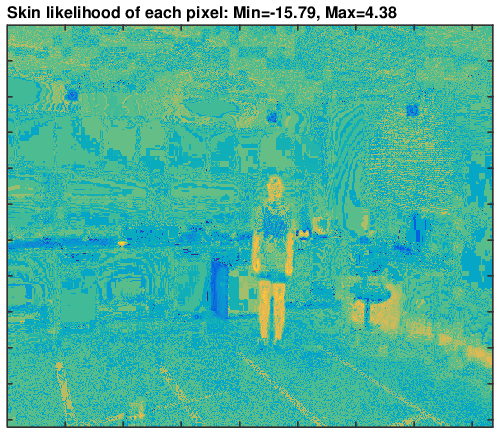}
 \includegraphics[width=0.11\textwidth]{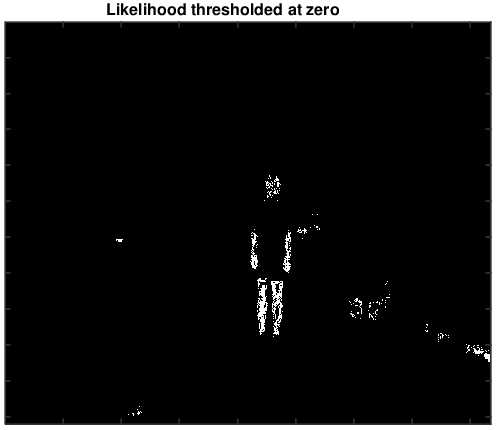} 
 \includegraphics[width=0.11\textwidth]{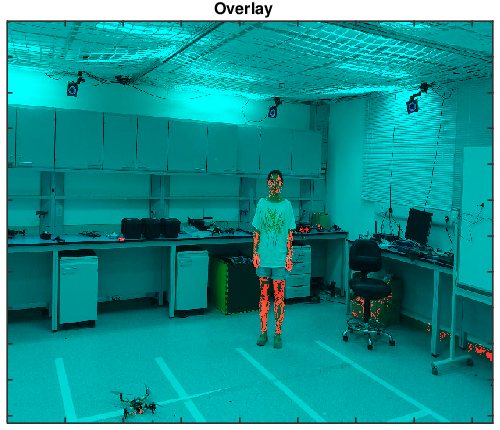} \\
 \includegraphics[width=0.11\textwidth]{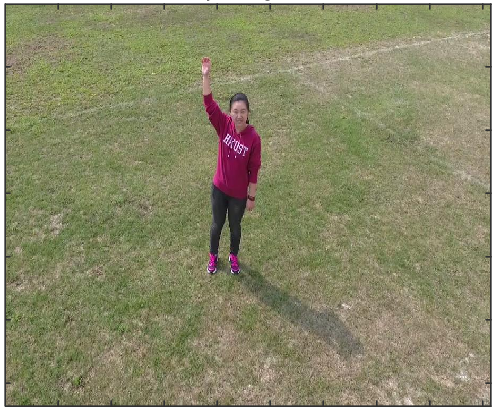} 
 \includegraphics[width=0.11\textwidth]{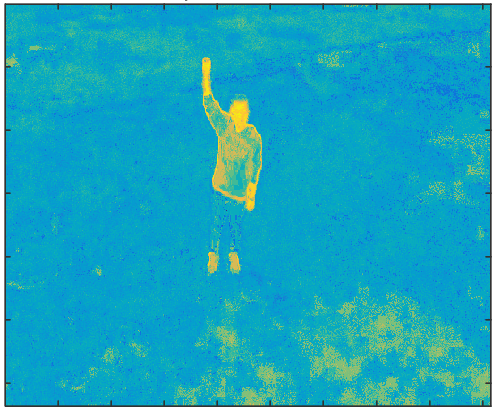}
 \includegraphics[width=0.11\textwidth]{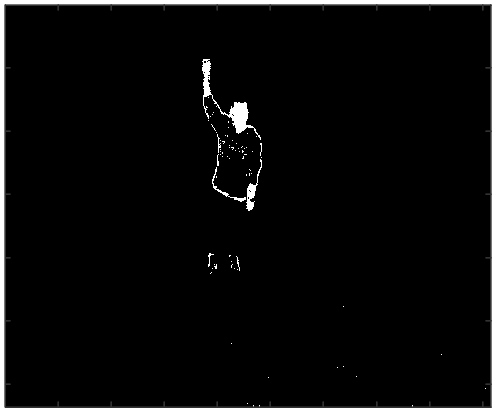}
 \includegraphics[width=0.11\textwidth]{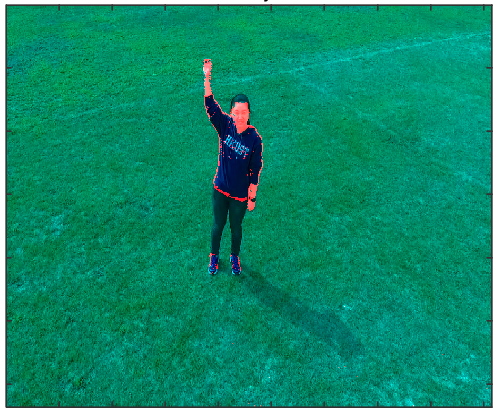} 
 \caption{Two examples of the skin detection \cite{conaire2007detector} results.   The first image of each row is the input image, the second one shows the skin likelihood of each pixel,  the third image shows the thresholded likelihood of each pixel (i.e, the binary skin image), and the last one shows the overlaid input image with the binary skin image through setting the R value of the skin pixels to 255.}
 \label{fig: skin detection}
 \vspace{-2em}
\end{figure}

In our system, we use the nonparametric histogram-based skin detection model in \cite{conaire2007detector} which is trained using 14,985,845 manually annotated skin pixels and 304,844,751 non-skin pixels.  Using this model to apply skin detection on a given image is simply a table-lookup, rearranging the three 8-bit color values of an RGB pixel into one number and then obtaining the corresponding probability that this pixel is a skin pixel.  Fig.~\ref{fig: skin detection} shows two examples of the skin detection results.  The first image of each row is the input image, the second one shows the skin likelihood of each pixel,  the third image shows the thresholded likelihood of each pixel (i.e, the binary skin image), and the last one shows the overlaid input image with the binary skin image through setting the R values of the skin pixels to 255.  In our system, the skin detection is applied in a rectangular region extended form the tracking bounding box, and we use the binary skin image as a mask and represent it as $\mathbf{M}_{skin}$.

Note that besides the hands and arms, the face, neck, legs and feet can also be detected as skin areas.  Assuming that the user is standing, we remove the skin areas within $\mathbf{B}_{user}$ roughly corresponding to these regions according to preset ratios, leaving an easier job for the hand detection.

\subsubsection{Hand Detection}
Here we use $\mathbf{p}$ to denote a 2D pixel coordinate in a image, $\mathcal{F}$ represents a function, $\mathbf{B}$ is a bounding box, $\mathbf{M}$ is a mask of the same size as the input frame and $\lambda$ is scaler weight.  Superscripts and subscripts are used to identify different variables of the same kind and also give information about their properties.  $\mathbf{p}_{\textit{uc}}$ is the center point between the shoulders of the user.  $\mathbf{s}_{\textit{hand}}$ is a set of position coordinates of the pixels in a square whose area is roughly equal to that of a hand.  We manually set the side length of this square to $\mathbf{B}_{user}.\textit{width} / 4$.

According to the gestures used in our interaction system, two kinds of hand positions can trigger the movement of the aerial robot: stretching out one hand parallel to the image plane and placing one hand in front of the body.  Since the first gesture corresponds to much richer commands, it is more often used and can be detected with higher confidence, the existence of a stretched out hand is checked first.  Whether there is a hand placed in front of the body is checked later if the detection of a stretched out hand fails.

Given $\mathbf{B}_{user}$ and $\mathbf{M}_{skin}$,  the stretched out hand is detected following the procedure in Algorithm~\ref{alg: stretching out hand detection}. We first check whether there is sufficient number of skin pixels outside of $\mathbf{B}_{user}$, which indicates the existence of a stretched out hand, and then the position of the hand is determined based on two considerations: the hand region should contain many skin pixels and the stretched out hand should have a large distance from $\mathbf{p}_{\textit{uc}}$.  The function $\mathcal{F}_{\textit{skin}}(\mathbf{p})$ simply counts the number of skin pixels in a $\mathbf{s}_{\textit{hand}}$ centered at $\mathbf{p}$.  From the formulation of $\mathcal{F}_{\textit{dist}}(\mathbf{p})$ it can be seen that we score the horizontal and vertical distances differently.  Thinking of the practical scene when a user is interacting with an aerial robot, in most cases a stretched out hand would be horizontally away from the center of the person, and since we do not distinguish between the left and right arm, the absolute value of the horizontal difference $|\mathbf{p}_{\textit{uc}}.x - \mathbf{p}.x|$ is used. However, when the user raises his/her arm high up, $|\mathbf{p}_{\textit{uc}}.x - \mathbf{p}.x|$ can be small, so $(\mathbf{p}_{\textit{uc}}.y - \mathbf{p}.y)$ is used to add a score onto the raised hand position.  The term $\mathcal{F}_{\textit{dist}}(\mathbf{p})$ not only captures the ``stretched" property, but also distinguishes the hand from the arm.  Since the hand is at the tip of the skin region, it is further away from the body than the arm when it is stretched out.

\begin{algorithm}[b]
 \caption{Stretched out hand detection}
\label{alg: stretching out hand detection}
\begin{algorithmic}[1]
\Require ~~\ 
$\mathbf{B}_{user}$, $\mathbf{M}_{skin}$
\Ensure ~~\ 
$\mathbf{p}_{\textit{hand}}^{\textit{out}}$

\State $\mathbf{M}_{\textit{skin}}^{\textit{out}} = \mathcal{F}_{\mathbf{B}_{user}}^{\textit{out}}(\mathbf{M}_{skin},\mathbf{B}_{user})$  \Comment {get skin region out of $\mathbf{B}_{user}$}
\If {number of non-zero elements in $\mathbf{M}_{\textit{skin}}^{\textit{out}} > 30$} 
\State $\mathbf{p}_{\textit{hand}}^{\textit{out}} = \underset{\mathbf{p} \in \mathbf{M}^{\textit{out}}}{\arg\max} (\mathcal{F}_{\textit{dist}}(\mathbf{p}) + \lambda_{1} \mathcal{F}_{\textit{skin}}(\mathbf{p}))$
\State where $\mathcal{F}_{\textit{dist}}(\mathbf{p}) = |\mathbf{p}_{\textit{uc}}.x - \mathbf{p}.x| + \lambda_{2} (\mathbf{p}_{\textit{uc}}.y - \mathbf{p}.y)$ 
\State and   $\mathcal{F}_{\textit{skin}}(\mathbf{p}) = \sum_{\mathbf{p'} \in \mathbf{s}_{\textit{hand}}^{\mathbf{p}}} \mathbf{M}_{\textit{skin}}^{\textit{out}} (\mathbf{p'})$  \Comment{count skin pixels}
\State $\mathbf{p}_{\textit{hand}}^{\textit{out}} = \mathbf{p}_{\textit{hand}}^{\textit{out}} - \mathbf{p}_{\textit{uc}}$  \Comment{hand position relative to the center point between shoulders} 
\Else 
\State $\mathbf{p}_{\textit{hand}}^{\textit{out}} = \mathbf{0}$  
\EndIf
\end{algorithmic}
\end{algorithm}

Similarly, the procedure to detect the hand in front of the body is described in Algorithm~\ref{alg: front body hand detection}, where $\mathbf{p}_{\textit{bc}}$ is the center of $\mathbf{B}_{user}$.  This time the hand is only searched for within $\mathbf{B}_{user}$.  An area with many skin pixels is again preferred, and the term $\mathcal{F}_{\textit{dist}}^{\textit{cost}}$ penalizes the positions that are horizontally away from the center of the body.

\begin{algorithm}
 \caption{Front-of-body hand detection}
\label{alg: front body hand detection}
\begin{algorithmic}[1]
\Require ~~\ 
$\mathbf{B}_{user}$, $\mathbf{M}_{skin}$
\Ensure ~~\ 
$\mathbf{p}_{\textit{hand}}^{\textit{front}}$

\State $\mathbf{M}_{\textit{skin}}^{\textit{in}} = \mathcal{F}_{\mathbf{B}_{user}}^{\textit{in}}(\mathbf{M}_{skin},\mathbf{B}_{user})$  \Comment{get skin region in $\mathbf{B}_{user}$}
\State $\mathbf{p}_{\textit{hand}}^{\textit{front}} = \underset{\mathbf{p} \in \mathbf{M}_{\textit{skin}}^{\textit{in}}}{\arg\max} (-\mathcal{F}_{\textit{dist}}^{\textit{cost}}(\mathbf{p}) + \lambda_{3} \mathcal{F}_{\textit{skin}}(\mathbf{p}))$
\State where $\mathcal{F}_{\textit{dist}}^{\textit{cost}}(\mathbf{p}) = |\mathbf{p}_{\textit{bc}}.x - \mathbf{p}.x|$
\State and   $\mathcal{F}_{\textit{skin}}(\mathbf{p}) = \sum_{\mathbf{p'} \in \mathbf{s}_{\textit{hand}}^{\mathbf{p}}} \mathbf{M}_{\textit{skin}}^{\textit{in}} (\mathbf{p'})$
\If {$\mathcal{F}_{\textit{dist}}^{\textit{cost}} < \mathbf{B}_{user}.\textit{width} / 5 \land \mathcal{F}_{\textit{skin}} > 30$}
\State $\mathbf{p}_{\textit{hand}}^{\textit{front}} = \mathbf{p}_{\textit{hand}}^{\textit{front}} - \mathbf{p}_{\textit{bc}}$
\Else 
\State $\mathbf{p}_{\textit{hand}}^{\textit{front}} = \mathbf{0}$
\EndIf
\end{algorithmic}
\end{algorithm}

\subsubsection{Command Generation}

Algorithm~\ref{alg: command generation} describes how we generate a command based on the results of previous steps.  The ``command" here simply means a 3D vector in the camera frame.  This vector indicates where the desired position is relative to the current position of the aerial robot.  The real target position for the control system is obtained by scaling this vector and transforming it into the body frame of the aerial robot.

We use $\mathbf{S}$ to denote the state buffer.  Specifically, $\mathbf{S}^{\textit{out}}$ stores all the $\mathbf{p}_{\textit{hand}}^{\textit{out}}$ detected in the latest $60$ frames, and $\mathbf{S}^{\textit{front}}$ stores the latest $60$ $\mathbf{p}_{\textit{hand}}^{\textit{front}}$. 

\begin{algorithm}
 \caption{Command generation}
\label{alg: command generation}
\begin{algorithmic}[1]
\Require ~~\ 
$\mathbf{S}^{\textit{out}}$, $\mathbf{S}^{\textit{front}}$
\Ensure ~~\ 
$\mathbf{p}_{\textit{cmd}}$

\State $n^{\textit{out}} \gets$ count non $\mathbf{0}$ in $\mathbf{S}^{\textit{out}}$  \Comment{check stretched out hand}
\If {$n^{\textit{out}} > 30$}  \Comment{a stretched out hand is detected in more than half of the frames in the state buffer}
\State $\mathbf{p}_{\textit{cmd}} \gets$ average of non $\mathbf{0}$ in $\mathbf{S}^{\textit{out}}$ \Comment{set two components of $\mathbf{p}_{\textit{cmd}}$, leaving $\mathbf{p}_{\textit{cmd}}.z = 0$}
\If {$|\mathbf{p}_{\textit{cmd}}.y| > |\lambda_{4} \mathbf{p}_{\textit{cmd}}.x|$}
\State $\mathbf{p}_{\textit{cmd}}.x = 0$
\EndIf

\Else  \Comment{check the front-of-body hand}
\State $n_{\textit{higher}} \gets$ count $\mathbf{p}_{\textit{hand}}^{\textit{front}}.y < 0$ in $\mathbf{S}^{\textit{front}}$
\State $n_{\textit{lower}} \gets$ count $\mathbf{p}_{\textit{hand}}^{\textit{front}}.y > 0$ in $\mathbf{S}^{\textit{front}}$
\If {$n_{\textit{higher}} > 30$}
\State $\mathbf{p}_{\textit{cmd}}.z = -1$   \Comment{come closer}
\ElsIf {$n_{\textit{lower}} > 30$}
\State $\mathbf{p}_{\textit{cmd}}.z = 1$   \Comment{go further}
\Else
\State $\mathbf{p}_{\textit{cmd}}.z = \mathbf{0}$
\EndIf
\EndIf
\end{algorithmic}
\end{algorithm}

As mentioned previously, the stretched out hand can be detected more confidently than the hand in front of the body, so, in general, the commands for the directions parallel to the image plane are more reliably recognized than those for directions perpendicular to the image plane.  Assuming that there is at most one true command each time, in our system a decision tree is used to generate the command.  The high-confidence choices are checked first, while the low-confidence choices are only checked when the previous attempt has failed.  

Generating a command based on a single frame is vulnerable to outliers. Our system first checks whether there are enough states in the buffer consistently containing non-resting gestures of either kind (stretching out one hand or placing a hand in front of the body).  If so, the recognition results of these frames are averaged to robustly generate a smooth piloting command.  This method increases the robustness and smoothness of the generated command, which is quite favorable from the practice and control points of view.  

Recall in Subsection~\ref{subsec:command design} that we mention adjusting our generated pointing up and down commands to make our system behave more reasonably.  From Algorithm~\ref{alg: command generation} it can be seen that this adjustment is done by simply setting the small $\mathbf{p}_{\textit{cmd}}.x$ to $0$.

\section{EXPERIMENTAL RESULTS}
\label{sec:experiments} 
We first test the capability of the face detection adopted in our system to automatically initialize the bounding box for DSST.  Then we test our recognition algorithm given the bounding box containing the user.  The performance of the entire system is shown last.  We implement our method on a DJI Matrice 100 with an Intel NUC computer and a Logitech webcam onboard.  We use ROS (Robot Operating System) to handle the communication between different modules.  Some of the video samples processed off-line to demonstrate the gesture design (in Subsection~\ref{subsec:command design}), face detection and skin detection (in Subsection~\ref{subsubsec:skin detection}) are taken by a DJI Phantom~4.  In our experiments we choose $\lambda_{1} = 0.5$, $\lambda_{2} = 0.2$, $\lambda_{3} = 0.013$ and $\lambda_{4} = 0.5$ empirically.

\subsection{Capability of Face Detection}
\label{subsec:capability of face detection}
\begin{figure*}
 \centering
  \vspace{0.15cm}
 \includegraphics[width=0.23\textwidth]{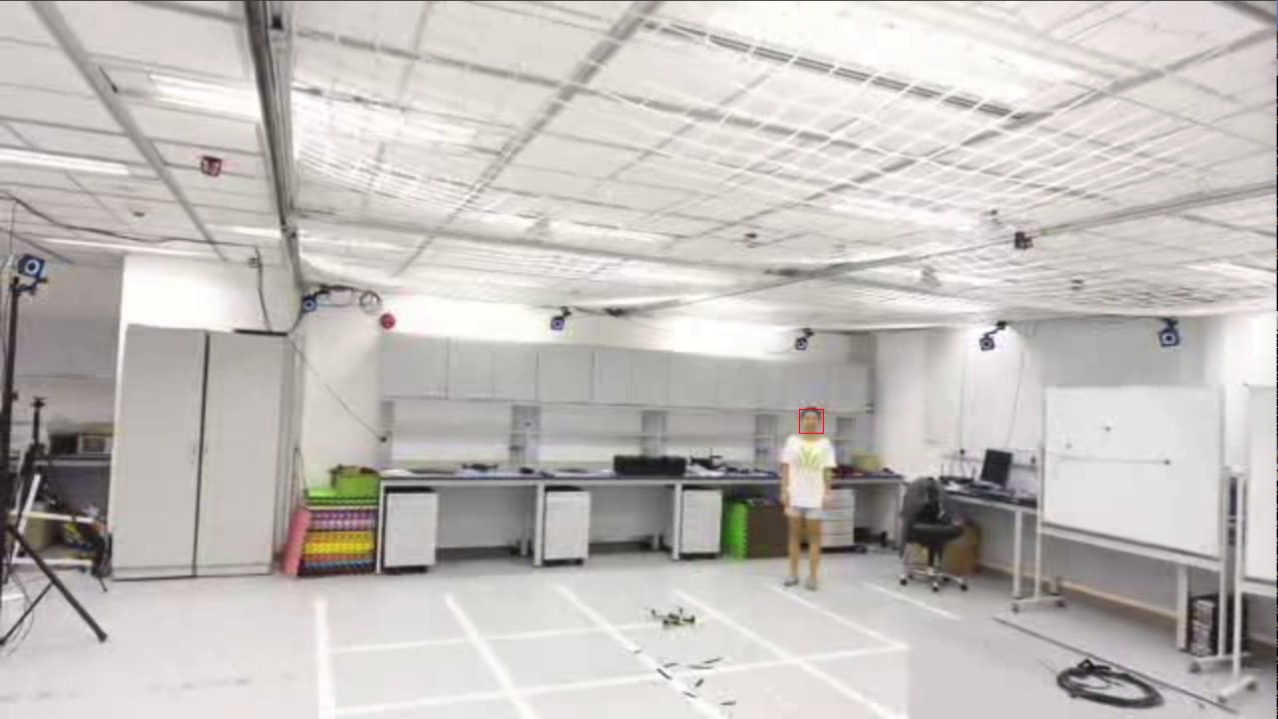}
 \includegraphics[width=0.23\textwidth]{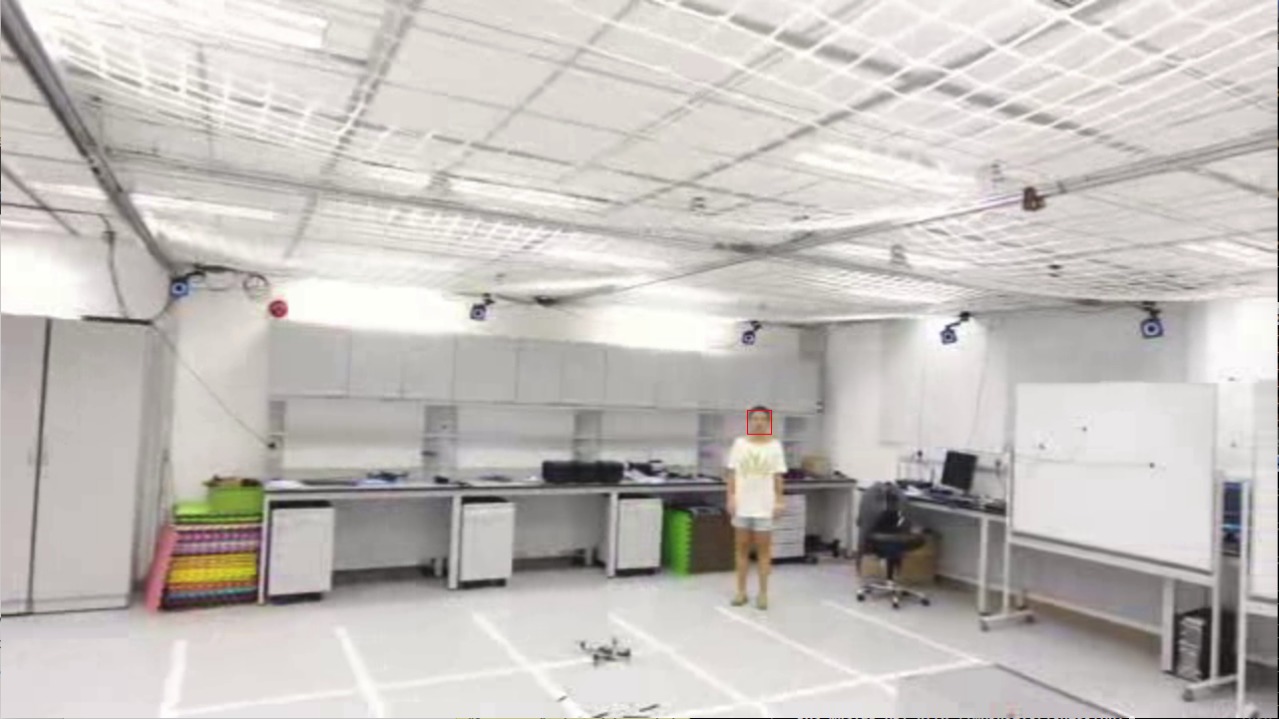} 
 \includegraphics[width=0.23\textwidth]{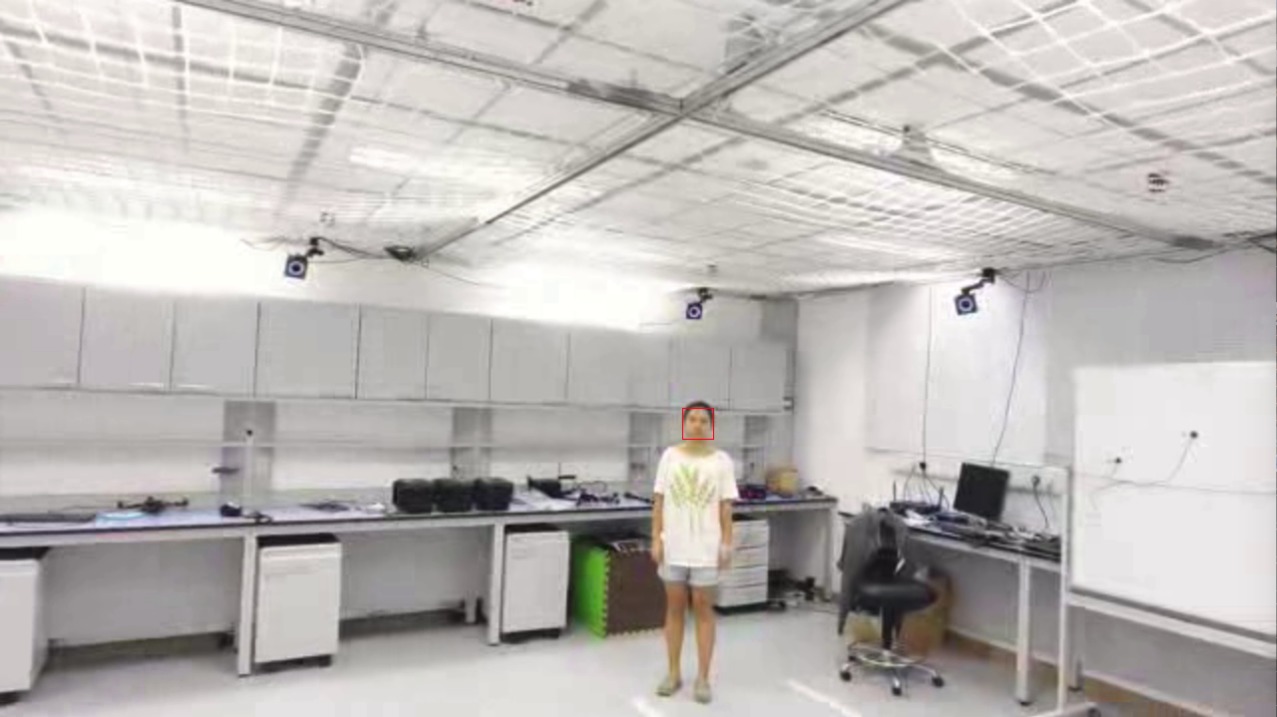} 
 \includegraphics[width=0.23\textwidth]{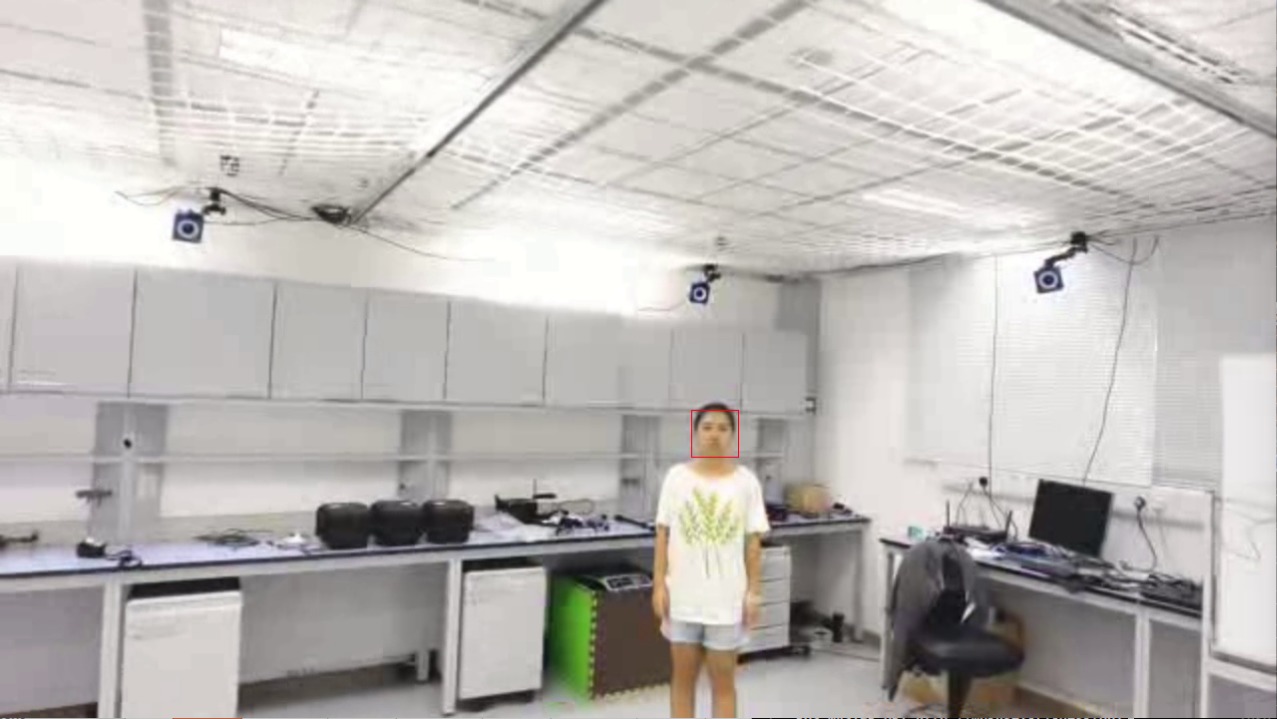} \\
 \includegraphics[width=0.3\textwidth]{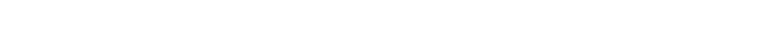} \\
 \includegraphics[width=0.23\textwidth]{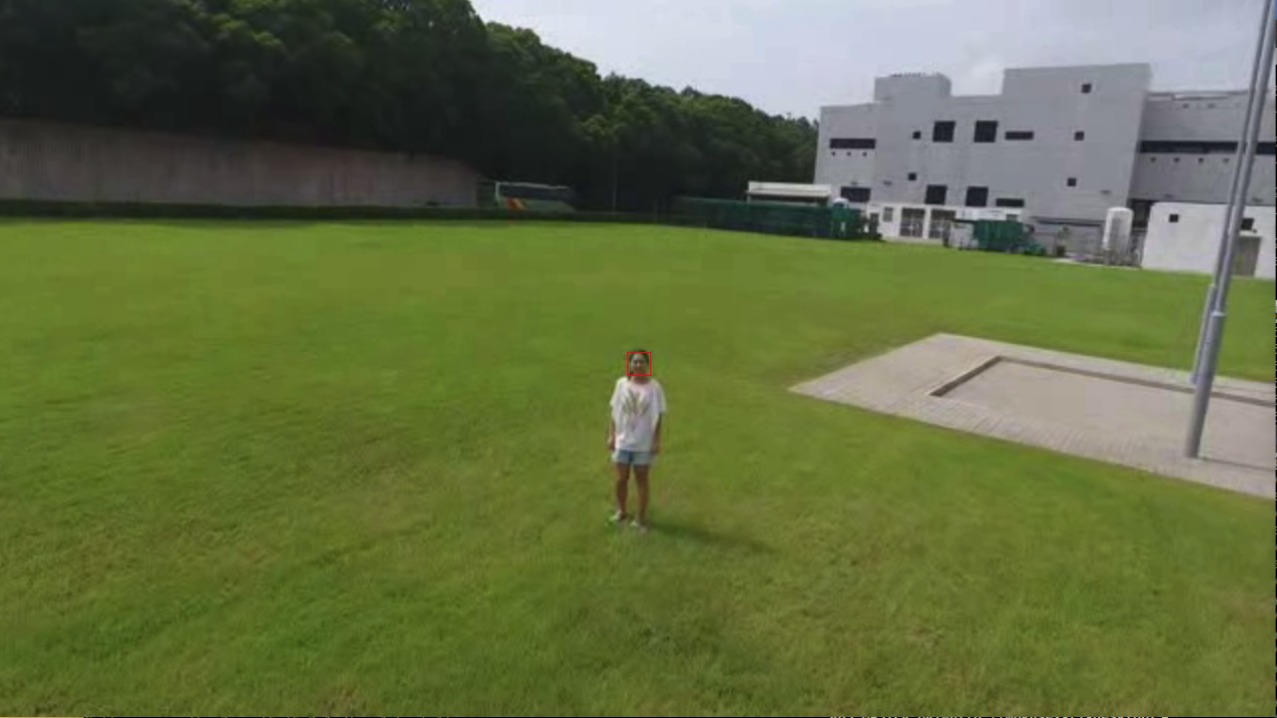} 
 \includegraphics[width=0.23\textwidth]{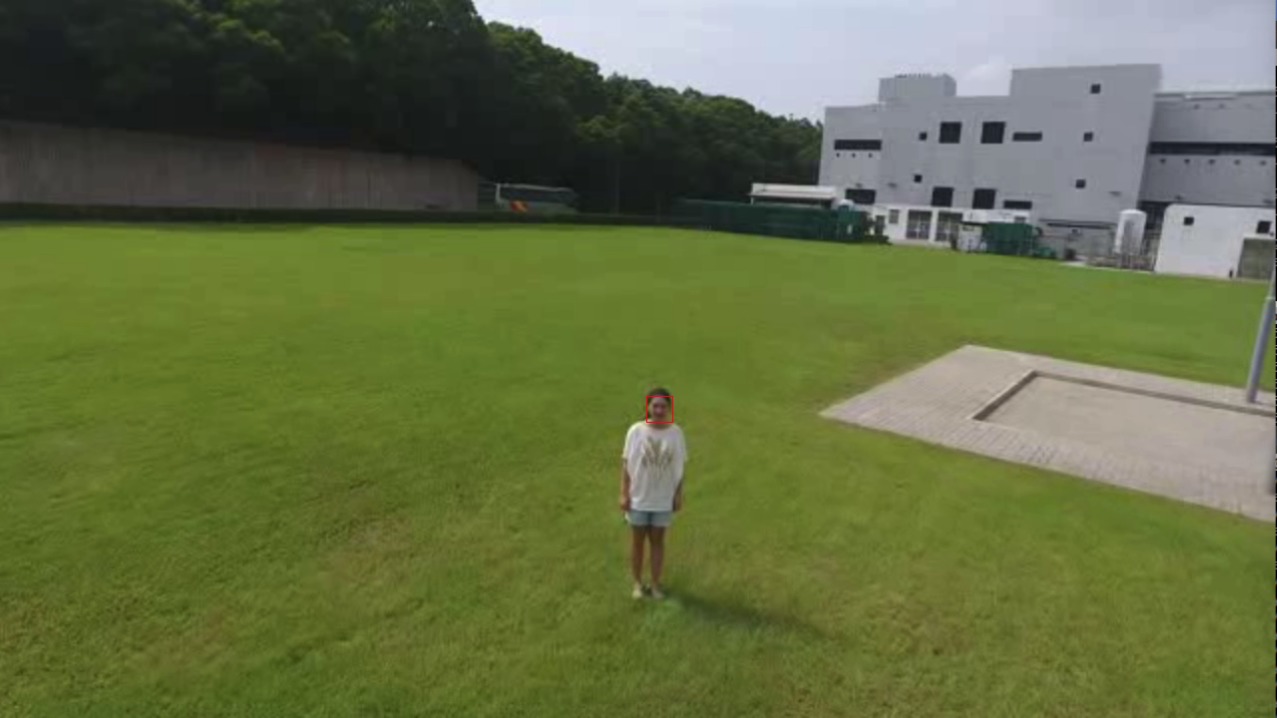}
 \includegraphics[width=0.23\textwidth]{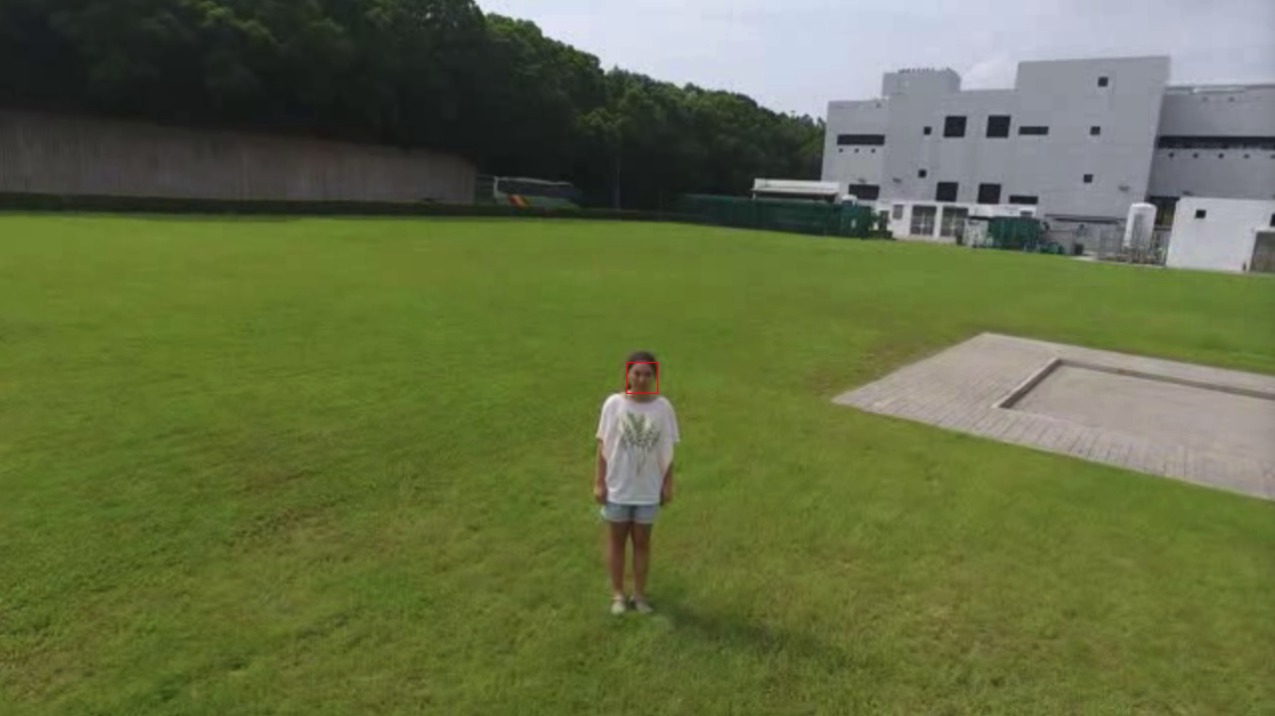}
 \includegraphics[width=0.23\textwidth]{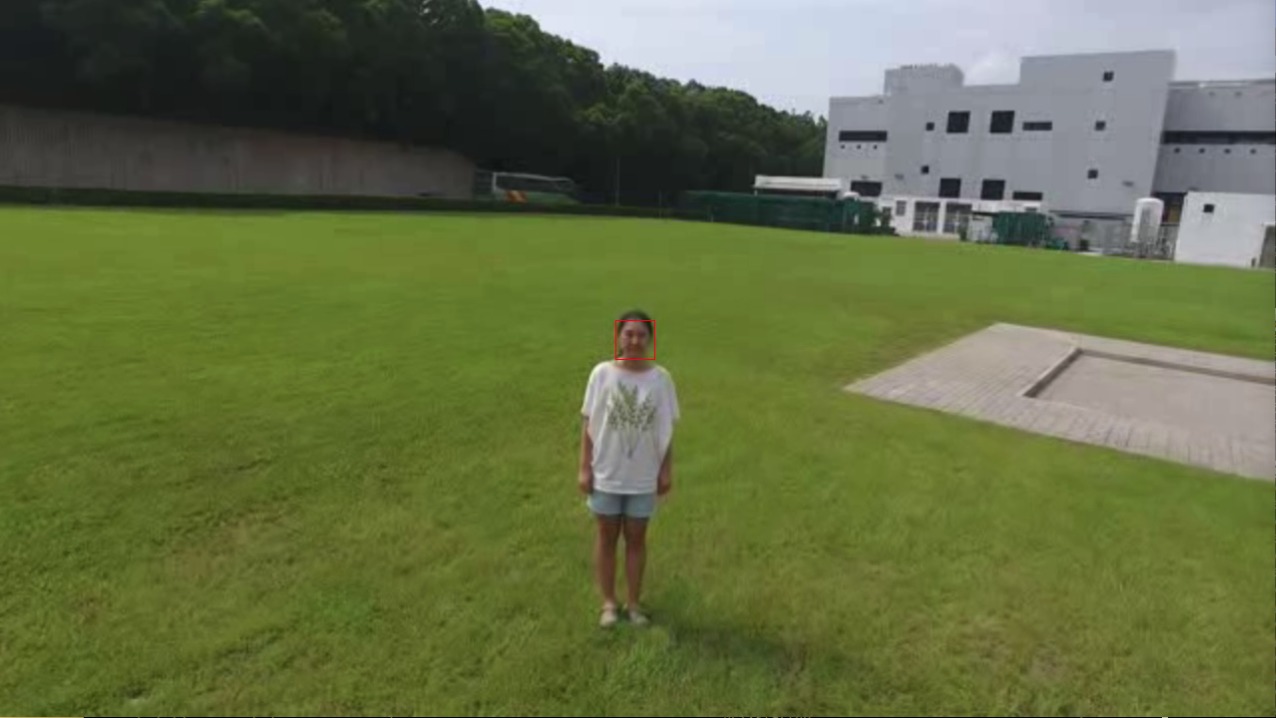}
 \caption{Face detection test on two 720p video sequences taken by an aerial robot moving towards a single person from far away.  The first row of frames is taken in an indoor environment, and the second one is outdoors.  From left to right the user takes up a larger and larger proportion of the image.  The face of the user is successfully detected in all eight frames and a red square bounding box tightly containing the face is also shown in each frame.  The left-most frame of each row shows the point at which the face detection just starts to work.}
 \label{fig: face detection}
  \vspace{-1em}
\end{figure*}

We test the capability of the face detection method by two 720p video sequences taken by the Phantom 4.  Both videos contain one person, and one video is taken indoors, while the other is taken outdoors.  Initially the quadrotor is far away from the person and the face detection fails to detect the face of the person. As the quadrotor gradually moves closer to the person, the face is more and more reliably detected.  Fig.~\ref{fig: face detection} shows some frames where face detection is successful.  The first row of frames is taken in an indoor environment and the second one is outdoors.  From left to right the person takes up a larger and larger proportion of the image.  The left-most image shows the point at which face detection starts to work.  Conservatively speaking, using our cameras (both the camera on the Phantom 4 and Logitech webcam), the face detection works within about $5$ meters of the person.  Note that this face detection is used to automatically initialize the tracker, so it only performs on the first frame of the video stream. I.e., $5$ meters is a conservative range within which the system can automatically initialize its tracker.  Since our tracker also has manual initialization implemented, it will work well even if the user is further away than $5$ meters when the system starts.  This distance also depends on the video quality, as higher quality videos enable the face detection to work at a longer distance.

\subsection{Performance of Gesture Recognition Algorithm}
The performance of our gesture recognition algorithm is shown in this subsection.  We first test the recognition accuracy of the stretched out hand gesture, and then show the recognition accuracy of placing a hand in front of the user.  Here we assume that the tracker has handled the camera motion and a reliable bounding box containing the user is given, so we simply use video taken by a stationary camera.

\begin{figure}
 \centering
 \includegraphics[width=0.49\textwidth]{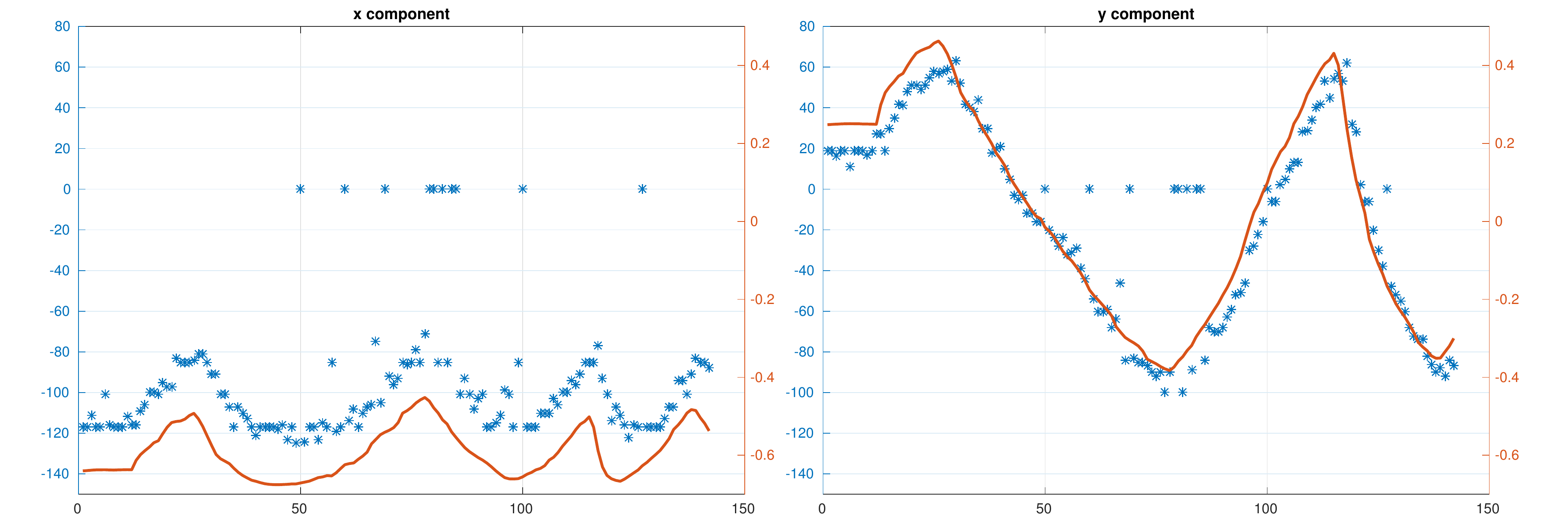}
 \caption{Hand detection accuracy in the image plane.  The orange plot is the ground truth obtained by the OptiTrack with its axis shown on the right, and the blue plot is the results obtained by our algorithm with its axis shown on the left.}
 \label{fig: hand detecton accuracy}
 \vspace{-1em}
\end{figure}

\begin{figure*}
 \centering
  \vspace{0.15cm}
 \includegraphics[width=0.137\textwidth]{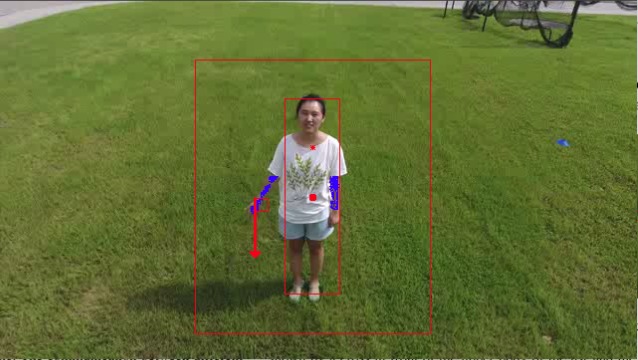}
 \includegraphics[width=0.137\textwidth]{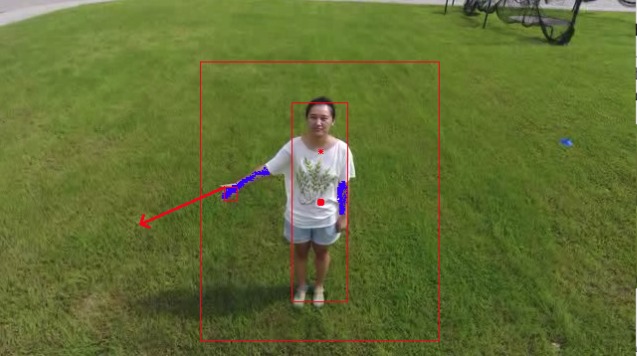}
 \includegraphics[width=0.137\textwidth]{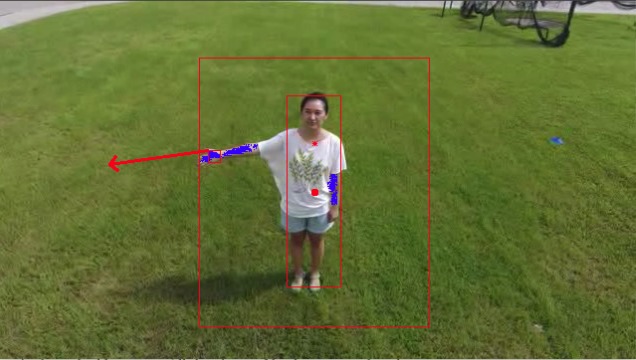}
 \includegraphics[width=0.137\textwidth]{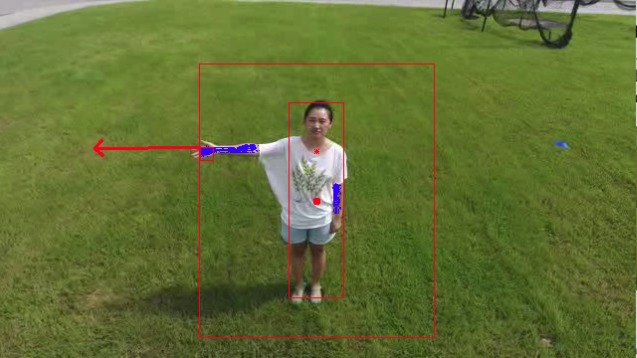}
 \includegraphics[width=0.137\textwidth]{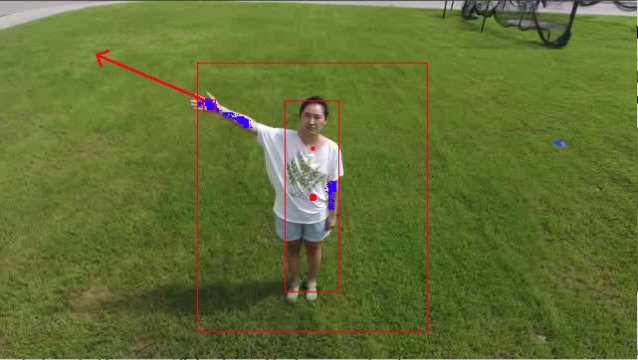}
 \includegraphics[width=0.137\textwidth]{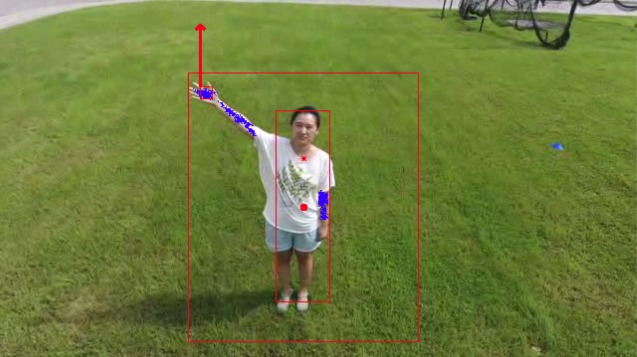}
 \includegraphics[width=0.137\textwidth]{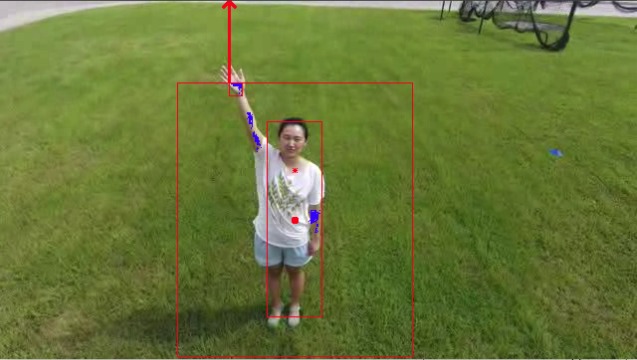}
 \caption{Sample frames of the video clip demonstrating fine piloting in the directions parallel to the image plane.}
 \label{fig: fine piloting}
  \vspace{-1em}
\end{figure*}

Since the directions pointed to by a stretched out hand of the user is directly used to pilot the aerial robot without quantization, the performance of this fine piloting is mainly determined by the accuracy of the recognized gesture vector, which starts from the center point between the shoulders of the user and ends at the position of the stretched out hand.  In order to test the recognition accuracy, we tie one marker to the middle of the shoulders of the user and one to the back of her stretched out hand.  An OptiTrack system is used to precisely record the 3D position coordinates of these two markers to obtain the ground truth gesture vector, and we only use two components of the 3D vector that are parallel to the image plane.  Fig.~\ref{fig: hand detecton accuracy} shows the detection accuracy of the $x$ and $y$ components w.r.t. the frame index.   Note that our recognition algorithm use the coordinate system of the image frame, which is different from that used by OptiTrack.  There is a translation between the two origins and their coordinates have different scales.  From Fig.~\ref{fig: hand detecton accuracy} it can be seen that for most of the frames, the recognition results of our algorithm are quite consistent with the ground truth.  In some frames our algorithm fails to detect the hand and the gesture vectors simply remain zero.  In the outdoor environment, we do not have the OptiTrack system to offer us a ground truth.  A processed video taken by a Phantom 4 hovering in front of the user is used to offer a subjective performance demonstration.  Both indoor and outdoor videos are included in our video attachment, from which it can be seen that even though the hand detection (represented by a red square box containing the hand, and also where the red arrow starts) fails occasionally, the generated command (the red arrow) is quite stable due to the state buffer adopted in our algorithm.  The outdoor video also well illustrates our fine piloting in the directions parallel to the image plane.   The user slowly raises her arm from its resting position and the generated command gradually changes accordingly.  Some representative frames of this outdoor video are shown in Fig.~\ref{fig: fine piloting}.

With the gesture of putting a hand in front of the body, either the ``come closer" or ``go further" command is triggered.  The system only needs to identify in which of the two regions (either pink or purple region in Fig.~\ref{fig:command design}) the hand is placed, rather than to determine the precise position of the hand.  A video clip in our attachment shows the recognition performance of our system on this pair of gestures.  We manually check the recognition result in each frame, and the success rate is $85.45\%$.

\subsection{Performance of the Entire System}
\label{subsec:Performance of the entire system}
We successfully implement the proposed system on a Matrice 100 with an Intel NUC and a Logitech webcam onboard.  The position of the quadrotor is controlled by the gesture of the user, and its yaw angel is automatically adjusted based on the position of the user relative to that of the quadrotor,  keeping the camera always facing the user.  A demonstration of mixed gesture piloting of the aerial robot is available in our video attachment.  For our hardware settings, the proposed system works within about $10$ meters.

We find seven labmates without particular knowledge of this system (but they may have some general knowledge of UAVs from the class) to evaluate our system.  We first demonstrate the gesture commands by interact with the quadrotor ourselves (it takes about $1.5$ minutes to go through all the gestures), then ask them to interact with the quadrotor for about $5$ minutes and to evaluate the system form four aspects: whether the gestures are comfortable and intuitive to perform; whether piloting an aerial robot through our system is enjoyable; the rough percentage of the correct response (accuracy); whether it is easy to make adjustment.  The accuracy is given by a percentage, and the other three evaluations score from $0$ to $9$.  $0$ means too bad or unacceptable, $9$ means excellent and $5$ means reasonably fine.  The averaged score and the standard deviation of each evaluation is shown in Table~\ref{tab:user evaluation}.  All the users think this gesture-based piloting system is enjoyable and the gestures are comfortable to perform in general.  Some users think that the commands of ``come closer" and ``go further" are not very intuitive,  and the percentage of failure cases of this pair of gestures are relatively high.  This is due to the imperfection of the skin detection and the design of our decision tree, which gives higher priority to the directions parallel to the image plane.  One user points out that the easiness of adjustment is hard to judge, since the motion of the quadrotor is not very precise anyway.  Another user sharply senses that there is a delay between when he performs the gesture and when the quadrotor starts to act correspondingly.  This is mainly due to two reasons: first, for safety reason, we can not allow the quadrotor to go too fast, so we manually insert a $600$ milliseconds delay before each gesture-command generated and this upper bounds the gesture-command to about $1.6667$ Hz; second, the usage of the state buffer, which is adopted to improve the robustness of the system and to smooth the command generated.  Here some other information of the system are offered: the speed of the control loop is about $103$ Hz, the frame rate of the received video is about $25$, the algorithm generates gesture-command with the average rate of $22.4$, and drifting is observed when the received GPS signal is weak.  Optimization of this system and even making it be able to adapt to each user would be our future work.

\begin{table}[h]
\begin{center}
{\small
\begin{tabular}{|l|c c c c|}
\hline
\multicolumn{1}{|c|}{Evaluation}    &  comfort  & enjoyment &  accuracy & easy adjust \\
\hline\hline
mean                      &  $7.7857$       & $7.7857$     & $0.7286$  & $7.2143$   \\
std             &  $0.9063$       & $0.3934$     & $0.0859$  &  $0.6362$ \\
\hline
\end{tabular}
}
\end{center}
\caption{The summary of the user evaluations.}
\label{tab:user evaluation}
\vspace{-1em}
\end{table}

\section{CONCLUSIONS}
\label{sec:conclusion}
In this paper we propose a monocular gesture-based piloting system for an aerial robot.  To command an aerial robot to fly in a direction parallel to the image plane, the user simply stretches out one arm and points in that direction.  This design is not only simple and intuitive, but also very convenient for the user to make fine adjustments.  Our system also recognizes a pair of gestures in which a hand is placed in front of the body to command the aerial robot to come closer or go further,  enabling piloting in 3D space.  A DSST is used to locate the user, and a face detector is adopted for automatic bounding box initialization.  Based on a skin mask result from a nonparametric histogram-based skin detection model together with the tracking result, the hand position in each frame is obtained.  The recognition results in multiple frames are combined together to robustly produce a command.  The entire system can successfully run on an onboard mini PC in real time.  Evaluations from seven users show that this gesture-based piloting system of an aerial robot is very enjoyable and intuitive to use.



\section*{ACKNOWLEDGMENT}
This work was supported by HKUST institutional studentship.  The author would like to thank the equipment support from DJI and our labmates Tianbo Liu, Fei Gao, Tong Qin, Ximin Lyu etc. for their help.

\bibliographystyle{IEEEtran} 
\bibliography{IEEEabrv,IEEEexample}

\end{document}